%% file: main.tex
\documentclass[sigconf]{acmart}

\usepackage{listings,multirow}
\usepackage{graphicx,rotating}
\usepackage{textcomp}
\usepackage{xcolor}
\usepackage{xspace}
\usepackage{subfigure}
\usepackage{subcaption}

\newif\iffinal

\iffinal
    \newcommand\joaquin[1]{}
    \newcommand\raj[1]{}
\else
    \newcommand\joaquin[1]{{\color{blue}[Joaquin: #1]}}
    \newcommand\raj[1]{{\color{purple}[Raj: #1]}}
\fi

\author{Flavio Castro, Weijian Zheng, Joaquin Chung, Ian Foster, and Rajkumar Kettimuthu}
\affiliation{
  \institution{Argonne National Laboratory}
  \city{ Lemont }
  \state{ IL }
  \country{USA}}
\email{{fcastro, wzheng, chungmiranda, foster, kettimut}@anl.gov}

\copyrightyear{2025} 
\acmYear{2025} 
\setcopyright{cc}
\setcctype{by}
\acmConference[SC Workshops '25]{Workshops of the International Conference for High Performance Computing, Networking, Storage and Analysis}{November 16--21, 2025}{St Louis, MO, USA}
\acmBooktitle{Workshops of the International Conference for High Performance Computing, Networking, Storage and Analysis (SC Workshops '25), November 16--21, 2025, St Louis, MO, USA}
\acmDOI{10.1145/3731599.3767455}
\acmISBN{979-8-4007-1871-7/2025/11}


\begin{document}

\title{To Stream or Not to Stream: Towards A Quantitative Model for Remote HPC Processing Decisions} 

\begin{abstract}
Modern scientific instruments generate data at rates that increasingly exceed local compute capabilities and, when paired with the staging and I/O overheads of file-based transfers, also render file-based use of  remote HPC resources impractical for time-sensitive analysis and experimental steering. Real-time streaming frameworks promise to reduce latency and improve system efficiency, but lack a principled way to assess their feasibility. In this work, we introduce a quantitative framework and an accompanying Streaming Speed Score to evaluate whether remote high-performance computing (HPC) resources can provide timely data processing compared to local alternatives. Our model incorporates key parameters including data generation rate, transfer efficiency, remote processing power, and file input/output overhead to compute total processing completion time ($T_{pct}$) and identify operational regimes where streaming is beneficial. We motivate our methodology with use cases from facilities such as APS, FRIB, LCLS-II, and the LHC, and validate our approach through an illustrative case study based on LCLS-II data.
Our measurements show that streaming can achieve up to 97\% lower end-to-end completion time than file-based methods under high data rates, while worst-case congestion can increase transfer times by over an order of magnitude, underscoring the importance of tail latency in streaming feasibility decisions.
\end{abstract}

\maketitle

\input{sections/intro}
\input{sections/motivation}
\input{sections/model}

\input{sections/approach}

\input{sections/evaluation}

\input{sections/conclusion}

\section*{Acknowledgment}
This work was supported in part by the Argonne Leadership Computing Facility and in part by the Diaspora project, both of which are supported by the U.S. Department of Energy under Contract DE-AC02-06CH11357.

\bibliographystyle{ACM-Reference-Format}
\bibliography{references}

\end{document}

%% file: sections/intro.tex
\section{Introduction} \label{sec:intro}

Modern scientific instruments generate data at rates that far exceed local processing capabilities. The Large Hadron Collider generates raw data at up to 40~TB/s, while SLAC's LCLS-II is expected to produce up to 1~Tbps. At Argonne's Advanced Photon Source (APS), detectors produce up to 480~Gb/s of data, rendering file-based post-processing inadequate. 
Real-time streaming analytics promises to reduce acquisition times by 44\%, enabling quality monitoring and experiment steering in order to increase experiment efficiency and reduce costly instrument downtime~\cite{thayer-2024,Vescovi-2022,tekin-2017}.

\begin{figure}[!t]                  
  \centering
  \includegraphics[width=\linewidth]{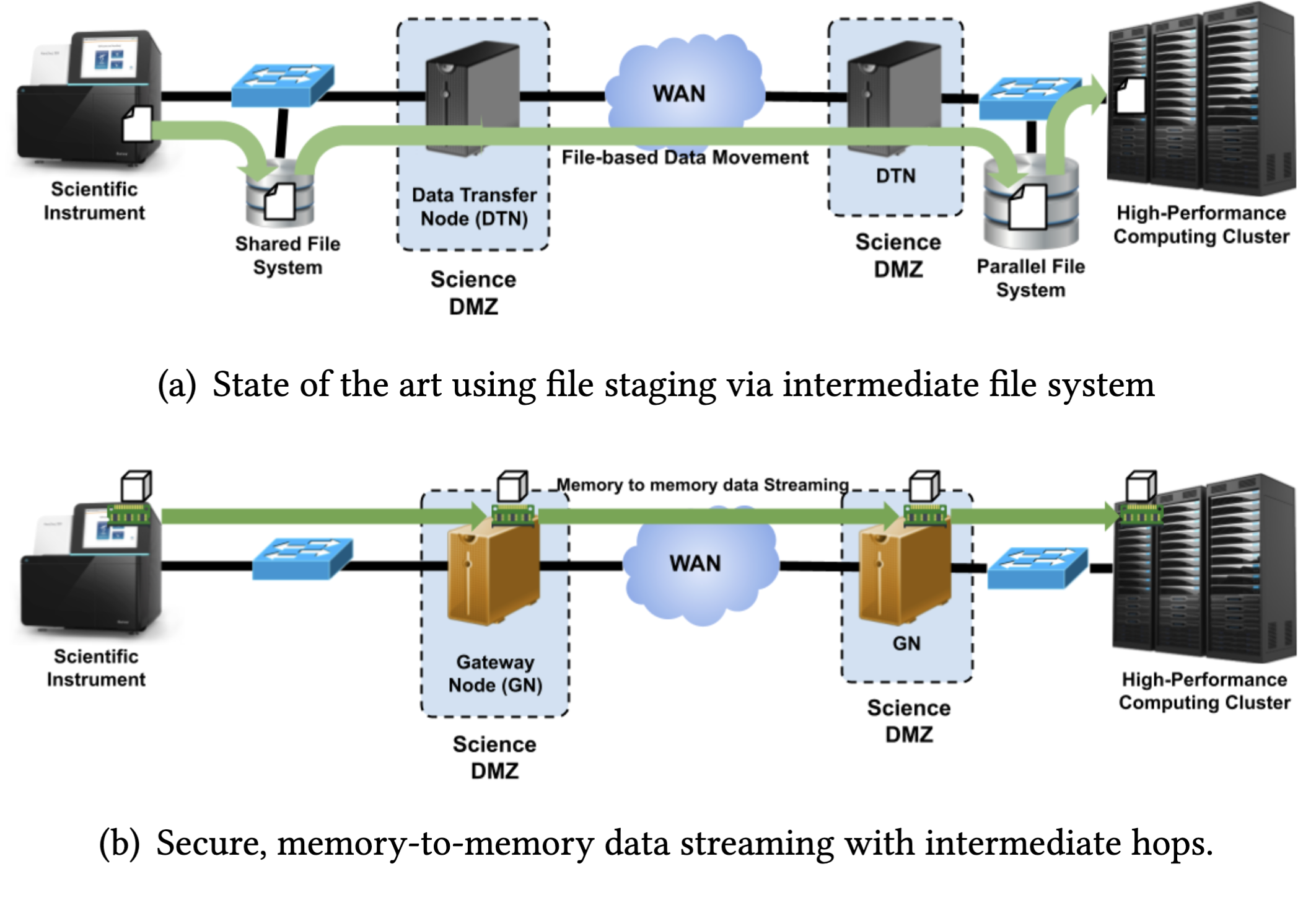}
  \caption{Data movement approaches for remote data analysis adapted from \cite{scistream-hpdc22}}
  \label{fig:filevsmemory}
  \vspace{-8.9ex}
\end{figure}

For remote data analysis, the prevailing approach (Figure~\ref{fig:filevsmemory}(a)) stages data via DTNs into the facility’s parallel file system, from which compute nodes then access it for processing. While this staging introduces disk I/O overhead, it can still deliver remarkable throughput for offline analysis, for example, achieving petabyte-per-day transfers~\cite{1pb-day}. However, such file-based workflows remain inadequate for near real-time analysis and experimental steering, where timely feedback is essential.

While recent streaming frameworks achieve 14$\times$ faster transfers than file-based methods~\cite{welborn-2024}, 
throughput alone cannot guarantee the subsecond latencies required for closed-loop autonomous control~\cite{sinisa-2023}.
Thus, streaming systems face significant network connectivity challenges (see Figure~\ref{fig:filevsmemory}(b)). 
The stochastic nature of network performance means that optimizing for maximum average throughput while ignoring tail latency leads to systematic failures in time-sensitive applications when faced with congestion. Previous performance evaluations have demonstrated that variation and average file transfer times of GridFTP v2 increase rapidly as traffic load increases~\cite{ohsaki-gridftp}.  We argue that tail latency is critical for experimental feedback loops, and systems that ignore outliers risk systematic failure.
Moreover, there are no consistent measurement frameworks to quantify these metrics in instrument-HPC systems, forcing ad-hoc design decisions.  Instrument facilities often resign to relying on HPC resources only for offline analysis, avoiding real-time processing requirements when local resources are not available.

We propose a quantitative framework for determining HPC streaming analysis feasibility that compares local versus remote processing. Our approach models data generation rates, processing complexity, flow completion time and I/O overhead, while providing a measurement framework for estimating streaming performance under congestion.
Our work makes the following contributions:
\begin{enumerate}
    \item A quantitative model assessing remote HPC streaming feasibility that incorporates data generation rates, network efficiency, processing ratios, and I/O overhead into a unified completion time ($T_{pct}$) metric.
    \item The Streaming Speed Score (SSS), a metric quantifying worst-case transfer time relative to ideal performance for evaluating network tail latency impact.
    \item A measurement methodology capturing worst-case flow completion times under controlled congestion to parameterize the model.
    \item A case study using APS and LCLS-II-inspired workloads demonstrating up to 97\% completion time reduction in high-rate scenarios and identifying regimes where local processing remains preferable.
\end{enumerate}

This paper is organized as follows: Section 2 critically assesses existing network performance measurement tools and examines state-of-the-art streaming applications. Section 3 presents our mathematical model for assessing local versus remote processing. Section 4 introduces our methodology for estimating streaming throughput and file transfer overhead. Section 5 applies our previous methodology in a hypothetical case study. Finally, Section 6 discusses the implications of our findings for different latency regimes and provides practical recommendations for facility decision-making.

%% file: sections/motivation.tex
\section{Motivation} \label{sec:motivation}
This section motivates our work from two angles. We first describe the limitations of current network performance measurement metrics and then highlight modern scientific streaming workflows with requirements that cannot be fulfilled by existing file-based data transfer methods.

\subsection{Network Performance Measurement Overview}
Current network measurement frameworks are largely focused on active probing methodologies and bulk-transfer optimization. The Fasterdata initiative of ESnet has established comprehensive benchmarks through standardized Data Transfer Node (DTN) configurations for achieving specific throughput thresholds at various  network scales. The Data Transfer Scorecard categorizes performance expectations by stakeholder perspective: researchers typically focus on data volume over time (TB/day), while network administrators examine instantaneous bitrates (Gbps) and link utilization percentages \cite{esnet-scorecard}. Active measurement approaches dominate current practice, with tools such as iPerf3, PerfSonar and GridFTP serving as primary instruments for characterizing network paths~\cite{swift22}.

Despite extensive optimization techniques, current methodologies exhibit systematic bias toward average performance that fundamentally misaligns with real-time analysis demands. These approaches treat congestion-free conditions as the normal case, dismissing congestion as outliers: a flawed assumption for networks without dedicated resource reservation. Traditional network research establishes that flow completion time (FCT) represents the most important user-centric metric, as it directly correlates with application-perceived performance \cite{dukkipati2006}. In addition, network tail latency is often a crucial metric for cloud application performance that can be affected by a wide variety of factors \cite{zhao2023}. This is especially important for applications with high loss sensitivity that tend to amplify impact of errors.

A critical class of streaming applications exhibits minimal resilience to network imperfections, requiring strict real-time completeness guarantees where neither message losses nor delays can be tolerated. These applications fundamentally differ from traditional bulk transfers in that incomplete data renders the entire computation invalid rather than merely degraded. This stringent completeness requirement transforms network reliability from a performance optimization concern into a fundamental correctness constraint.

The tomography reconstruction pipeline described in~\cite{tekin-2017} is a perfect example of a scientific application with high sensitivity to losses.
Missing a single projection degrades the whole process since reconstructing three-dimensional structures requires projections from all angles.
Similarly, the DELERIA streaming framework is also not tolerant of message losses, where dropped packets can cascade into processing pipeline failures \cite{ornl}. In these scenarios, worst-case latency dominates system design considerations instead of average throughput metrics.

To illustrate this distinction, let us consider an instrument generating 1~MB frames at 1~kHz, aggregating data for quality monitoring computations every second. Assume that quality monitoring can only be executed once all frames have been received. If the worst-case flow completion time for all frames generated in a second is 5 seconds, then the workflow experiences a minimum 5-second delay regardless of average network performance. This delay directly impacts experimental steering decisions, where scientists must wait transfer time plus remote processing time ($T_{transfer} + T_{remote}$) before receiving feedback. Consequently, we suggest that real-time streaming frameworks should incorporate explicit loss tolerance mechanisms and prioritize worst-case performance bounds over average-case optimization.
These measurement biases become critical when we examine how actual scientific facilities operate, where the ``outlier'' congestion events that current tools dismiss can invalidate hours of experimental time and millions of dollars in beam allocation.
\subsection{Science Drivers} \label{sec:science_drivers}

The following prominent use cases from major scientific facilities demonstrate the critical need for real-time data streaming architectures. These are examples of the application types that have driven the evolution from traditional file-based workflows to streaming-based approaches.

\subsubsection{Large Hadron Collider Experiments}
At the Large Hadron Collider, trigger strategies extract relevant information from vast streams of data in real time. Experiments such as ATLAS, CMS, and LHCb face extreme data challenges with proton-proton collisions occurring at 40 MHz, generating raw data rates up to 40~TB/s that must be reduced to approximately 1~GB/s for permanent storage. ATLAS and CMS, for example, utilize two-tier systems with low-level hardware-based triggers reducing rates from 40~MHz to 100~kHz within 4~$\mu$s, followed by software-based High-Level Triggers that further reduce output to 1~kHz. These experiments leverage heterogeneous computing architectures including FPGAs, GPUs, and CPU farms to perform real-time event reconstruction, alignment, calibration, and physics object identification, enabling analysis that would be impossible without such aggressive online data reduction strategies~\cite{lhc2024}.

\subsubsection{The Linac Coherent Light Source II}
LCLS-II at the SLAC National Accelerator Laboratory presents data rates scaling from 200~GB/s in 2023 to more than 1~TB/s in 2029. The facility streams data from ultra-high repetition rate imaging detectors (up to 1~MHz) through an adaptable Data Reduction Pipeline (DRP) that reduces data volume by an order of magnitude using experiment-specific algorithms including lossless compression, feature extraction, and software triggering. The system provides real-time feedback within 1 second for experimental steering and data quality monitoring within 1-10 minutes through the Analysis Monitoring Interface (AMI) framework. For computationally intensive experiments requiring exascale resources, LCLS-II automatically streams data to NERSC leadership computing facilities over ESnet, with the ExaFEL project demonstrating the ability to reduce serial femtosecond crystallography analysis time from weeks to minutes, processing TB/s data streams for real-time molecular structure determination~\cite{thayer-2024}.

\subsubsection{Advanced Photon Source (APS) Real-Time Tomographic Reconstruction}
At APS, real-time tomographic reconstruction streaming is demonstrated with data rates reaching 10s of GB/s. The system streams X-ray projection data from APS beamlines to the Argonne Leadership Computing Facility (ALCF) using a distributed stream processing architecture with up to 1,200 cores. The system achieves reconstruction rates up to 204 projections per second with less than 3\% overhead, enabling real-time feedback loops for autonomous experimental steering~\cite{tekin-2017}.

\subsubsection{DELERIA}
This project streams gamma-ray detector data from the Facility for Rare Isotope Beams (FRIB) to remote HPC systems at 40~Gbps (targeting 100~Gbps) over ESnet. DELERIA leverages nanomsg for network communication and containerized processes for data analysis, demonstrating the feasibility of remote, high-bandwidth streaming for time-sensitive nuclear physics experiments. Over 100 parallel analysis processes perform signal decomposition on waveforms, producing a 240~MB/s event stream for real-time monitoring, roughly 2~MB/s per compute process. This architecture enables scientists to optimize detector configurations between experimental runs, achieving a data reduction of 97.5\% while preserving all physics-relevant information~\cite{ornl}.

We observe a few constants across these use cases: extremely high data generation rates that make storage impractical, necessitating efficient data reduction strategies. Increasingly complex quality monitoring and offline analysis workflows that favor the usage of remote HPC versus extensive investments in local compute at each instrument facility. Notice that these requirements are ever-evolving and while we focus on bringing up examples from the state-of-the-art streaming use cases we predict that these patterns will become more common throughout instrument facilities once the streaming challenges are addressed.  

%% file: sections/model.tex
\section{Mathematical Model: Remote vs. Local Processing}
\label{sec:model}

In this section, we outline a decision framework to determine when streaming data provides advantages over data transfers and when remote HPC resources provide advantages over local processing. Our model focuses on total processing completion time ($T_{pct}$) as the primary metric for comparison. 

Previous analyses~\cite{BITTENCOURT2025100782} have characterized packet delays building upon the decomposition of Kurose and Ross~\cite{kurose2025computer}: processing delay, queuing delay, transmission delay, and propagation delay.
\begin{figure*}[t!]
    \centering
    \subfigure[]{\label{fig:m500_results}{\includegraphics[width=0.48\textwidth]{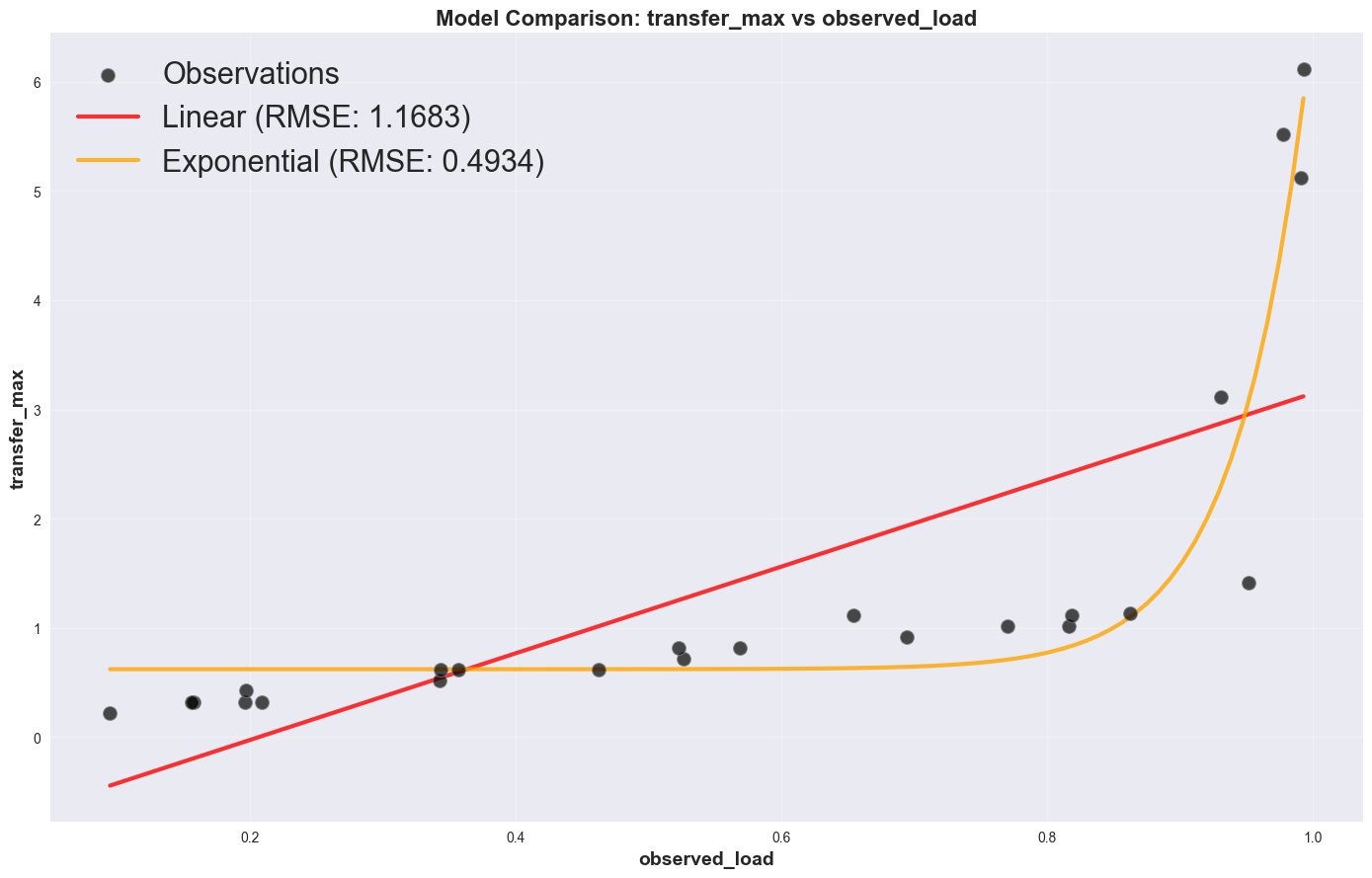}}}
    \subfigure[]{\label{fig:m500v2_results}{\includegraphics[width=0.48\textwidth]{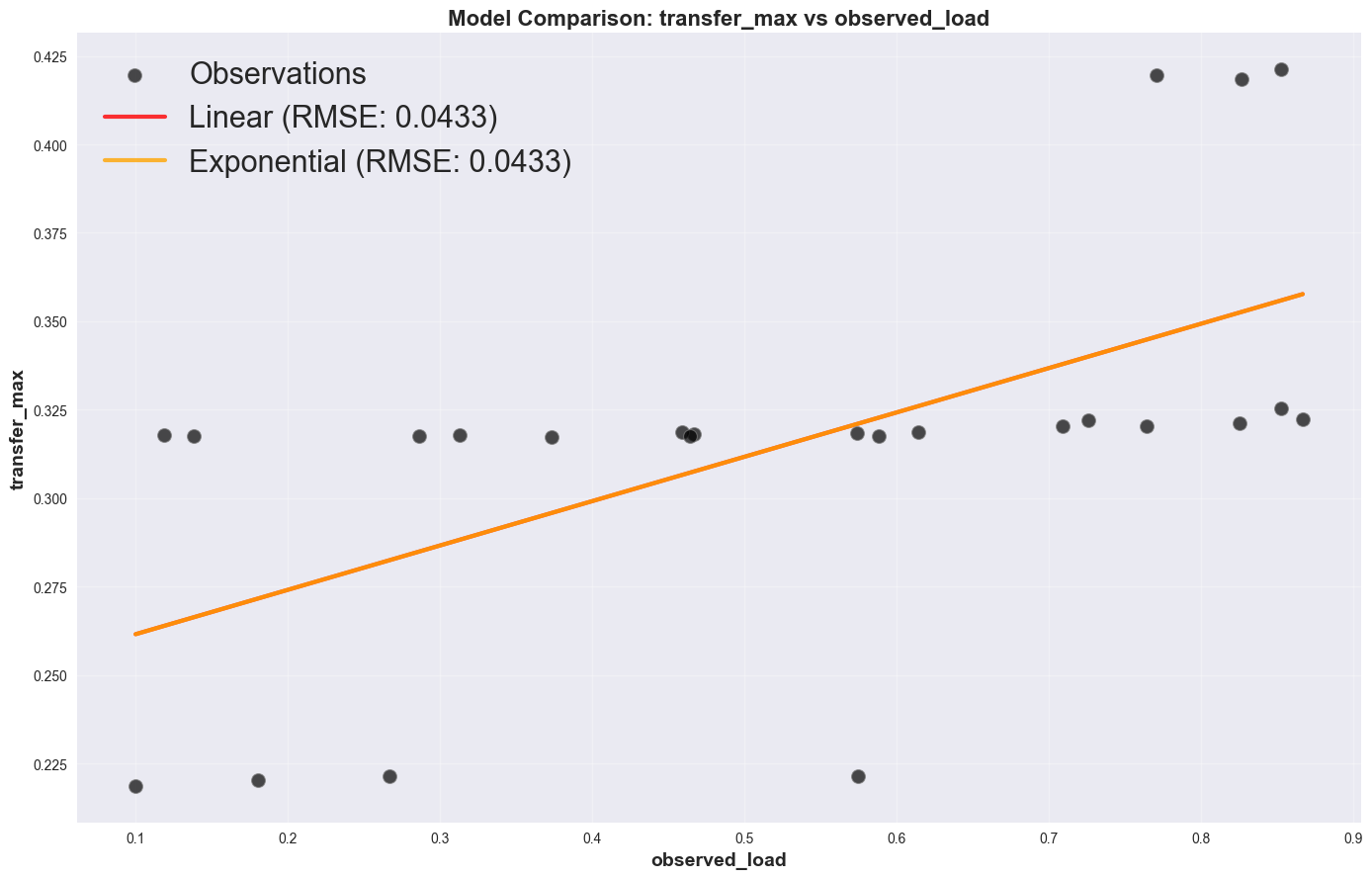}}}
    \caption{Maximum transfer time vs load for 0.5~GB transfers with
    \textit{P} = 2, 4, and 8 parallel TCP flows: (a) Simultaneous batches show non-linear growth of transfer time
    above 90 \% utilization due to congestion
    (b) Scheduled batches maintain steady transfer.}
    \label{fig:results}
    \vspace{-2.0ex}
\end{figure*}

\begin{equation}
d_{total} = d_{proc} + d_{queue} + d_{trans} + d_{prop}
\end{equation}

The authors of~\cite{BITTENCOURT2025100782} then simplify their model by assuming that as network capacity increases, processing and queuing delays approach zero while transmission delays become negligible due to increased bandwidth, leaving only propagation delays:

\begin{equation}
d_{continuum} \approx d_{prop}
\end{equation}

However, while this represents an elegant simplification, it falls precisely into the trap we warned about in our motivation section. The condition required for replacing flow completion time with propagation delay is that queuing delays must be absolutely zero, which would also imply zero packet loss. This represents an optimal-case analysis that fails to prepare for the outliers and congestion events that dominate real-time streaming failures. Moreover, this analysis implicitly assumes UDP-like transmission, ignoring the additional complexities introduced by TCP's congestion control, retransmissions and flow control mechanisms. 

Instead of ignoring queues and losses we argue for embracing some complexity. Our model incorporates not only data transfer overheads but also file I/O overheads. Rather than focusing on individual packet delays, we attempt to model total transfer time, capturing the actual application-perceived performance including the impact of congestion, retransmissions, and protocol overheads. Our approach recognizes that worst-case performance -- not average-case optimization -- drives the feasibility of time-sensitive scientific workflows.

\subsection{Parameters}

Our model employs the following parameters:
\begin{itemize}
    \item $S_{unit}$: Data unit size (GB)
    \item $C$: Computation complexity coefficient (FLOP/GB)
    \item $R_{local}$: Local processing rate (TFLOPS)
    \item $R_{remote}$: Remote processing rate (TFLOPS)
    \item $Bw$: Bandwidth (GBps)
    \item $R_{transfer}$: Effective data transfer rate (GB/s)
    \item $\alpha = R_{transfer}/Bw$: Transfer efficiency coefficient
    \item $r = R_{remote}/R_{local}$: Remote processing coefficient
    \item $\theta$: I/O overhead coefficient capturing file transfer overhead
\end{itemize}

\subsection{Basic Model}

For local processing, the total processing completion time is simply:
\begin{equation}
T_{local} = \frac{C \cdot S_{unit}}{R_{local}}
\end{equation}

For remote processing, we model the total processing completion time as:
\begin{equation}
T_{pct} = T_{transfer} + T_{remote} + T_{IO}
\end{equation}

where:
\begin{align}
T_{transfer} &= \frac{S_{unit}}{R_{transfer}} = \frac{S_{unit}}{\alpha \cdot Bw} \\
T_{remote} &= \frac{C \cdot S_{unit}}{R_{remote}} = \frac{C \cdot S_{unit}}{r \cdot R_{local}}
\end{align}

Since we define the I/O overhead relationship as:
\begin{align}
\theta &= \frac{T_{IO} + T_{transfer}}{T_{transfer}} \\
\theta \cdot T_{transfer} &= T_{IO} + T_{transfer}
\end{align}

Thus:
\begin{equation}
T_{pct} = \theta \cdot T_{transfer} + T_{remote}
\end{equation}

where $\alpha$ captures transfer efficiency (effective transfer rate over available bandwidth), $r$ is the ratio of remote to local processing speed, and $\theta$ accounts for file I/O overhead. Therefore:
\begin{equation}
T_{pct} = \frac{\theta \cdot S_{unit}}{\alpha \cdot Bw} + \frac{C \cdot S_{unit}}{r \cdot R_{local}}
\end{equation}







%% file: sections/approach.tex
\section{Experimental Evaluation} \label{sec:approach}


In this section, we estimate the file transfer overhead and total transfer time in illustrative scenarios to characterize remote analysis overhead. Note this is not a comprehensive analysis and rather an application of our current methodology to a few real scenarios. In future work we intend to explore this in more realistic comprehensive scenarios.

We define the following experiment that emulates worst-case congestion on data streaming performance to validate our framework. Our experiments characterize $T_{transfer}$ under varying levels of network load, providing empirical evidence and measurement of network behavior representative of congestion scenarios induced by TCP production real-time streaming scientific workflows.

The experimental orchestrator spawns multiple iperf3 server instances across sequential ports. Notice there is no contention on the server side, which is an optimistic requirement. The client component generates controlled network load by spawning iperf3 clients at a specified concurrency rate (clients per second), with each client transferring a configurable data volume using parallel TCP flows. The FABRIC~\cite{fabric-2019} server specs are presented in the Table~\ref{tab:hw}.

\begin{table}[!t]
\centering
\caption{Experimental Testbed Configuration}
\label{tab:hw}
\begin{tabular}{|l|l|}
\hline
\textbf{Component} & \textbf{Specification} \\
\hline
CPU & AMD EPYC 7532 (16 vCPUs) \\
Memory & 32 GB RAM \\
Network Interface & Mellanox ConnectX-5 (25 Gbps) \\
MTU & 9000 bytes (jumbo frames) \\
OS & Ubuntu 22.04.5 LTS \\
Kernel & Linux 5.15.0-143 \\
Virtualization & KVM \\
\hline
\end{tabular}
\vspace{-3.0ex}
\end{table}
Note that we implement two client spawning strategies: simultaneous batch spawning that creates instantaneous congestion spikes, and scheduled spawning between clients to model scheduled transfers. The  scripts used in this work are publicly available~\cite{github}.

Table~\ref{tab:parameters} summarizes the experimental configuration for the first experiment series. The experiments systematically vary concurrency and parallel flows while maintaining a fixed 0.5~GB transfer size, typical of instrument data frames. Each experiment ran 10 seconds of repeated transfers (e.g., 8 x 0.5 GB/second for 80 total transfers).

We collect network-level metrics (interface byte/packet counters) and application-level performance indicators (detailed transfer time logs per client). The maximum transfer time within each experiment serves as a heuristic for estimating worst-case performance under measured network utilization, providing practical insights into when streaming to remote HPC resources remains viable despite congestion.

\subsection{Streaming Speed Measurement Methodology}

 \begin{figure}[!t]                  
   \centering
   \includegraphics[width=\linewidth]{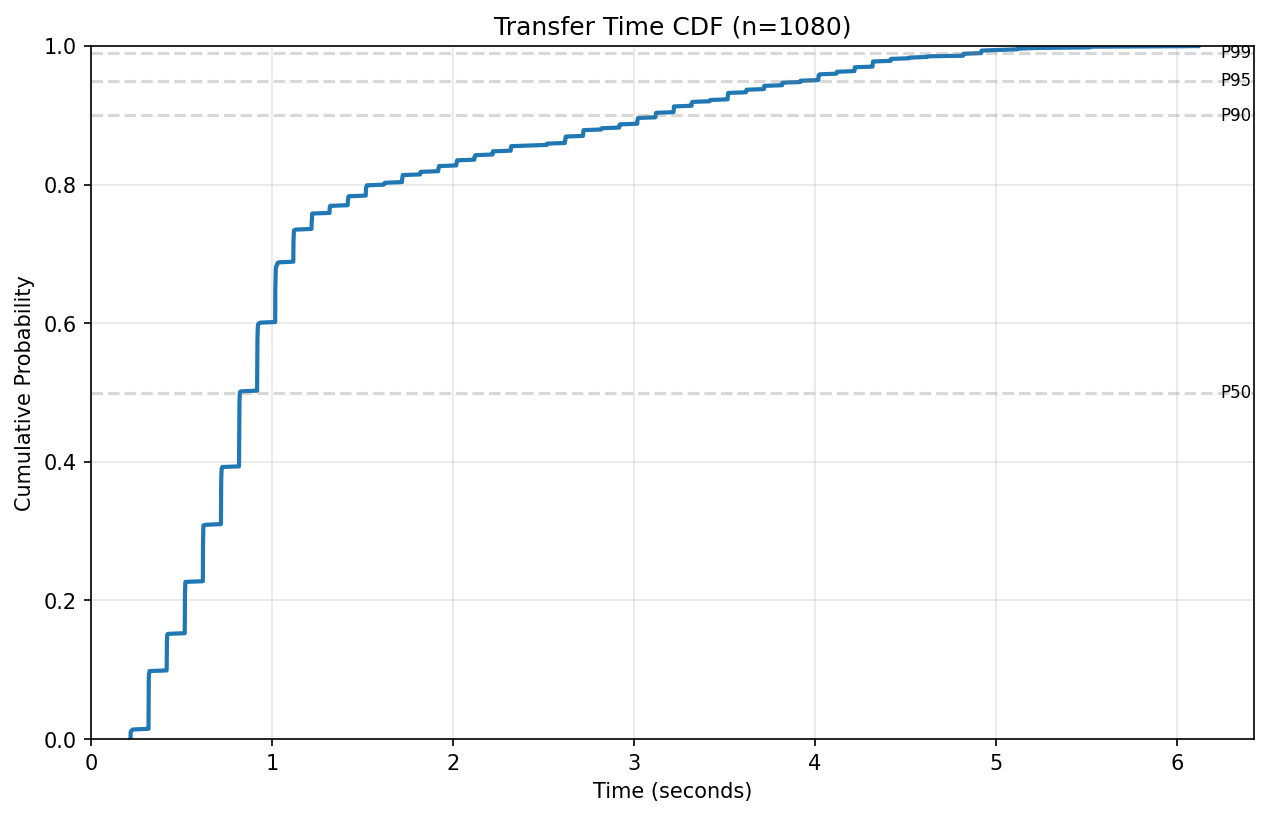}
   \caption{Cumulative probability distribution of Total transfer time including each file transfer. }
   \label{fig:time_cdf}
   \vspace{-1.0ex}
 \end{figure}

\begin{table}[!t] 
\centering
\caption{Experimental Configuration}
\label{tab:parameters}
\begin{tabular}{|l|c|l|}
\hline
\textbf{Parameter} & \textbf{Value/Range} & \textbf{Description} \\
\hline
Duration & 10~s & Experiment duration \\
Concurrency & 1--8 & Simultaneous clients \\
Parallel flows & 2, 4, 8 & TCP flows per client \\
Transfer size & 0.5~GB & Data volume per client \\
Total experiments & 24 & Full parameter sweep \\
Network interface & 25~Gbps & Mellanox ConnectX-5 \\
Round Trip Time & 16~ms & Ping results \\
\hline
\end{tabular}
\vspace{-3.0ex}
\end{table}

We define the Streaming Speed Score as the ratio of worst-case $T_{transfer}$ to theoretical minimum transfer time. 
\begin{align}
SSS = T_{worst} / T_{theoretical}, 
\end{align}
where $T_{worst}$ is the maximum observed transfer time under congestion and $T_{theoretical}$ is the ideal transfer time computed as data size divided by link bandwidth, which represents only the transmission delay component of the total delay. To measure the score's inputs, we conduct controlled congestion experiments by spawning concurrent data transfers at varying load levels, recording the maximum completion time across all transfers as $T_{worst}$.

Figure~\ref{fig:m500_results} illustrates the worst case transfer times in the presence of congestion. While theoretical transfer time for 0.5~GB at 25~Gbps is 0.16 seconds, observed maximum transfer times exceed five seconds at high utilization. The results clearly delineate three operational regimes: (1) low congestion with performance suitable for real-time applications, (2) moderate congestion with 2-3 second transfer times, and (3) severe congestion where transfer times become much higher and unsuitable for time-sensitive analysis. These empirical measurements provide illustrative examples for calculating streaming feasibility across different application latency requirements.

Figure~\ref{fig:m500v2_results} illustrates optimal transfer times when every transfer is scheduled to a specific time slot, and network bandwidth is reserved. The measured transfer time is ~0.2s -- within the error margin of the 0.16s theoretical value -- and the maximum transfer time remains comfortably within the 1-second time budget.
As shown in Figure~\ref{fig:time_cdf}, the total transfer time exhibits long-tail behavior, with non-linear increases at the P90 and P99 levels. We argue that these worst-case latencies should dominate system design considerations.

\subsection{File Transfer Overhead}

Our quantitative framework, grounded in the Streaming Speed Score and total processing completion time ($T_{pct}$) model, provides a way to assess this trade-off by accounting for data rates, transfer efficiency, and file I/O overhead. The following results illustrate these considerations using a representative Advanced Photon Source (APS) to Argonne Leadership Computing Facility (ALCF) transfer scenario.

\begin{figure}[!t]
    \centering
    \includegraphics[width=0.35\textwidth, height=0.135\textwidth]{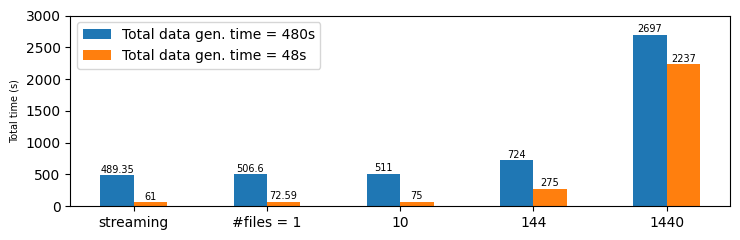}
    \caption{Comparison of streaming and file-based data transfer performance between the APS Voyager GPFS file system and the ALCF Eagle Lustre file system.}
    \label{fig:streaming_vs_file}
    \vspace{-3.0ex}
\end{figure}


Figure~\ref{fig:streaming_vs_file} compares memory-based streaming and file-based transfers between the APS Voyager GPFS and ALCF Eagle Lustre file systems for two data generation rates (0.033 s/frame and 0.33 s/frame). The scenario simulates transferring a single scan from an APS experimental facility: 1,440 frames of 2048 × 2048 pixels, totaling approximately 12.6 GB when stored as 2-byte unsigned integers.
At high frame rates, streaming minimized transfer time by overlapping transmission with generation, enabling near-immediate availability for remote processing. File-based workflows incurred increasing delays as file count grew, with the 1,440 small-file case suffering severe penalties from aggregation and metadata overhead. Even partial aggregation (e.g., 10 or 144 files) introduced noticeable delays.
Streaming's continuous flow eliminated aggregation waits and consistently outperformed file-based transfers at high frame rates. While file-based methods remain competitive at lower data rates or with large aggregated files, they cannot match streaming's immediacy for real-time analysis and experimental steering.

%% file: sections/evaluation.tex
\section{Case study} \label{sec:evaluation}

In this section, we aim to apply the model described in the previous sections to real use case requirements adapted from \cite{thayer-2024}. 
To analyze the feasibility of a scientific streaming-analysis workflow
under three tiers of $T_{pct}$:

\begin{itemize}
\item Tier 1 (Real-time analysis): $<$ 1s $T_{pct}$
\item Tier 2 (Near real-time analysis): $<$ 10s $T_{pct}$
\item Tier 3 (Quasi real-time analysis): $<$ 1min. $T_{pct}$
\end{itemize}

Using the workflows from Table~\ref{tab:lcls-workflows} as demonstrative cases, we evaluated whether remote HPC processing could meet Tier 1 and Tier 2 real-time operational deadlines while performing the offline analysis.  Notice that we extrapolate the measurements from Figure~\ref{fig:m500_results} to estimate the worst-case total transfer time for these use cases and disregard file transfer overheads.

First, let us explore the coherent scattering workflow that requires 2~GB/s sustained transfer rates to support remote analysis and requires an estimated compute power of 34~TF for analysis. Based on a 64\% utilization, we estimate the worst-case data streaming time to be 1.2 seconds. It would be well within the time constraints for Tier 2, while still leaving 8.8 seconds for the analysis ($T_{remote}$). If the instrument facility has the capacity to perform the analysis locally within less than 1.2 seconds, then local processing is favored. 

Now, let us examine the Liquid Scattering workflow that demands 4~GB/s sustained transfer rates and requires 20~TF of compute power. Obviously 4~GB/s (32~Gbps) would be unfeasible because it is higher than our link capacity of 25~Gbps. In this scenario the real-time capabilities are limited by local processing power.

For continuity of the analysis, we assume that we could further reduce transfer rates to 3~GB/s (24~Gbps). Based on a 96\% utilization we estimate the worst-case data streaming time to be 6 seconds. This would reduce the time constraints, leaving only 4 seconds for the remote analysis.

\begin{table}[!t] 
\centering
\caption{Compute-intensive workflows at LCLS-II showing throughput after 10x data reduction and computational requirements for 2023 (adapted from~\cite{thayer-2024})}
\label{tab:lcls-workflows}
\begin{tabular}{l|p{0.2\columnwidth}|p{0.15\columnwidth}}
\hline
\textbf{Description} & \textbf{Throughput} & \textbf{Offline Analysis} \\
\hline
Coherent Scattering (XPCS, XSVS) & 2 GB/s & 34 TF \\
\hline
Liquid Scattering & 4 GB/s & 20 TF \\
\hline
\end{tabular}
\vspace{-4.0ex}
\end{table}

%% file: sections/conclusion.tex
\section{Conclusion} \label{sec:conclusion}
We presented a quantitative framework to guide decisions on when remote HPC streaming can outperform local or file-based processing for time-sensitive scientific workflows. Our model captures key trade-offs between network transfer efficiency, remote compute capacity, and file I/O overhead through a gain function based on three core parameters: $\alpha$ (transfer efficiency), $r$ (remote-to-local processing ratio), and $\theta$ (I/O overhead). We motivated our approach drawing on use cases from facilities such as APS, FRIB, LCLS-II, and the LHC, and illustrated it with a case study inspired by LCLS-II data. Preliminary results show that streaming can reduce total processing completion time by over 97\% in representative scenarios, significantly improving data availability for downstream analysis. Future work will extend the model to incorporate concurrency, queuing effects, and variability in network and compute performance, as well as validate the framework across a broader range of facilities and configurations.
